\documentclass[twocolumn]{aastex631}

\usepackage{amsmath}

\begin{document}

\title{First rest-frame infrared spectrum of a $z>7$ quasar: JWST/MRS observations of J1120+0641}

\correspondingauthor{Sarah E.~I.~Bosman}
\email{bosman@mpia.de}

\author[0000-0001-8582-7012]{Sarah E.~I.~Bosman}
\affiliation{Max-Planck-Institut f\"{u}r Astronomie, K\"{o}nigstuhl 17, 69117 Heidelberg, Germany}
\affiliation{Institute for Theoretical Physics, Heidelberg University, Philosophenweg 12, D--69120, Heidelberg, Germany}

\author[0000-0002-7093-1877]{Javier \'{A}lvarez-M\'{a}rquez}
\affiliation{Centro de Astrobiolog\'{i}a (CAB), CSIC-INTA, Ctra. de Ajalvir km 4, Torrej \'{o}n de Ardoz, E-28850, Madrid, Spain}

\author[0000-0002-9090-4227]{Luis Colina}
\affiliation{Centro de Astrobiolog\'{i}a (CAB), CSIC-INTA, Ctra. de Ajalvir km 4, Torrej \'{o}n de Ardoz, E-28850, Madrid, Spain}

\author[0000-0003-4793-7880]{Fabian Walter}
\affiliation{Max-Planck-Institut f\"{u}r Astronomie, K\"{o}nigstuhl 17, 69117 Heidelberg, Germany}

\author[0000-0001-6794-2519]{Almudena Alonso-Herrero}
\affiliation{Centro de Astrobiolog\'{i}a (CAB), CSIC-INTA, Camino Bajo del Castillo s/n, E$-$28692 Villanueva de la Ca\~{n}ada, Madrid, Spain}

\author[0000-0003-1810-0889]{Martin J.~Ward}
\affiliation{Centre for Extragalactic Astronomy, Department of Physics, Durham University, South Road, Durham DH1 3LE, United Kingdom}

\author[0000-0002-3005-1349]{G\"{o}ran \"{O}stlin}
\affiliation{Department of Astronomy, Stockholm University, Oscar Klein Centre, AlbaNova University Centre, 106 91 Stockholm, Sweden}

\author[0000-0002-2554-1837]{Thomas R.~Greve}
\affiliation{Cosmic Dawn Center (DAWN), Denmark}
\affiliation{DTU Space, Technical University of Denmark, Elektrovej, Building 328, 2800, Kgs.~Lyngby, Denmark}
\affiliation{Dept. of Physics and Astronomy, University College London, Gower Street, London WC1E 6BT, United Kingdom}

\author[0000-0001-7416-7936]{Gillian Wright}
\affiliation{UK Astronomy Technology Centre, Royal Observatory Edinburgh, Blackford Hill, Edinburgh EH9 3HJ, United Kingdom}


\author[0000-0001-8068-0891]{Arjan Bik}
\affiliation{Department of Astronomy, Stockholm University, Oscar Klein Centre, AlbaNova University Centre, 106 91 Stockholm, Sweden}

\author[0000-0002-3952-8588]{Leindert Boogaard}
\affiliation{Max-Planck-Institut f\"{u}r Astronomie, K\"{o}nigstuhl 17, 69117 Heidelberg, Germany}

\author[0000-0001-8183-1460]{Karina I.~Caputi}
\affiliation{Kapteyn Astronomical Institute, University of Groningen, P.~O.~Box 800, 9700 AV Groningen, The Netherlands}

\author[0000-0001-6820-0015]{Luca Costantin}
\affiliation{Centro de Astrobiolog\'{i}a (CAB), CSIC-INTA, Ctra. de Ajalvir km 4, Torrej \'{o}n de Ardoz, E-28850, Madrid, Spain}

\author[0000-0001-6049-3132]{Andreas Eckart}
\affiliation{Physikalisches Institut der Universit\"{a}t zu K\"{o}ln, Z\"{u}lpicher Str.~77, 50937 K\"{o}ln, Germany}

\author[0000-0003-4801-0489]{Macarena Garc\'{i}a-Mar\'{i}n}
\affiliation{European Space Agency, Space Telescope Science Institute, Baltimore, Maryland, USA}

\author[0000-0001-9885-4589]{Steven Gillman}
\affiliation{Cosmic Dawn Center (DAWN), Denmark}
\affiliation{DTU Space, Technical University of Denmark, Elektrovej, Building 328, 2800, Kgs.~Lyngby, Denmark}

\author[0000-0001-9818-0588]{Manuel G\"{u}del}
\affiliation{University of Vienna, Department of Astrophysics, Türkenschanzstrasse 17, 1180 Vienna, Austria}
\affiliation{Max-Planck-Institut f\"{u}r Astronomie, K\"{o}nigstuhl 17, 69117 Heidelberg, Germany}
\affiliation{Institute of Particle Physics and Astrophysics, ETH Zurich,
Wolfgang-Pauli-Str 27, 8093 Zurich, Switzerland}

\author[0000-0002-1493-300X]{Thomas Henning}
\affiliation{Max-Planck-Institut f\"{u}r Astronomie, K\"{o}nigstuhl 17, 69117 Heidelberg, Germany}

\author[0000-0002-4571-2306]{Jens Hjorth}
\affiliation{DARK, Niels Bohr Institute, University of Copenhagen, Jagtvej 128, 2200 Copenhagen, Denmark}

\author[0000-0001-8386-3546]{Edoardo Iani}
\affiliation{Kapteyn Astronomical Institute, University of Groningen, P.~O.~Box 800, 9700 AV Groningen, The Netherlands}

\author[0000-0002-7303-4397]{Olivier Ilbert}
\affiliation{Aix Marseille Université, CNRS, LAM (Laboratoire d'Astrophysique de Marseille) UMR 7326, 13388, Marseille,
France}

\author[0000-0002-2624-1641]{Iris Jermann}
\affiliation{DTU Space, Technical University of Denmark, Elektrovej, Building 328, 2800, Kgs.~Lyngby, Denmark}
\affiliation{Cosmic Dawn Center (DAWN), Denmark}

\author[0000-0002-0690-8824]{Alvaro Labiano}
\affiliation{Telespazio UK for the European Space Agency, ESAC, Camino Bajo del Castillo s/n, 28692 Villanueva de la Ca\~nada, Spain}

\author{Pierre-Olivier Lagage}
\affiliation{AIM, CEA, CNRS, Universit\'{e} Paris-Saclay, Universit\'{e} Paris
Diderot, Sorbonne Paris Cit\'{e}, F-91191 Gif-sur-Yvette, France}

\author[0000-0001-5710-8395]{Danial Langeroodi}
\affiliation{DARK, Niels Bohr Institute, University of Copenhagen, Jagtvej 128, 2200 Copenhagen, Denmark}

\author[0000-0002-9850-2708]{Florian Pei{\ss}ker}
\affiliation{Physikalisches Institut der Universit\"{a}t zu K\"{o}ln, Z\"{u}lpicher Str.~77, 50937 K\"{o}ln, Germany}

\author[0000-0002-2110-1068]{Tom P.~Ray}
\affiliation{Dublin Institute for Advanced Studies, Astronomy \& Astrophysics Section, 31 Fitzwilliam Place, Dublin 2, Ireland}

\author[0000-0002-5104-8245]{Pierluigi Rinaldi}
\affiliation{Kapteyn Astronomical Institute, University of Groningen, P.~O.~Box 800, 9700 AV Groningen, The Netherlands}

\author{Martin Topinka}
\affiliation{Dublin Institute for Advanced Studies, Astronomy \& Astrophysics Section, 31 Fitzwilliam Place, Dublin 2, Ireland}

\author[0000-0001-7591-1907]{Ewine F.~van Dishoeck}
\affiliation{Leiden Observatory, Leiden University, PO Box 9513, Leiden, The Netherlands}

\author[0000-0001-5434-5942]{Paul van der Werf}
\affiliation{Leiden Observatory, Leiden University, PO Box 9513, Leiden, The Netherlands}

\author[0000-0002-1368-3109]{Bart Vandenbussche}
\affiliation{Institute of Astronomy, KU Leuven, Celestijnenlaan 200D bus 2401,
3001 Leuven, Belgium}


\begin{abstract}

We present a JWST/MRS spectrum of the quasar J1120+0641 at $z=7.0848$, the first spectroscopic observation of a reionisation-era quasar in the rest-frame infrared wavelengths ($0.6 < \lambda_\text{rest} < 3.4 \mu$m). 
In the context of the mysterious fast assembly of the first supermassive black holes at $z>7$, our observations enable for the first time the detection of hot torus dust, the H$\alpha$ emission line, and the Paschen-series broad emission lines in a quasar at $z>7$, which we compare to samples at $z<6$ for signs of evolution. 
Hot torus dust is clearly detected as an upturn in the continuum emission at $\lambda_{\text{rest}}\simeq1.3\mu$m, leading to a black-body temperature of $T_\text{dust} = 1413.5_{-7.4}^{+5.7}$ K. Compared to similarly-luminous quasars at $0<z<6$, the hot dust in J1120+0641 is somewhat elevated in temperature (top $1\%$). The temperature is more typical among $6<z<6.5$ quasars (top $25\%$), leading us to postulate a weak evolution in the hot dust temperature at $z>6$ ($2\sigma$ significance).
We measure the black hole mass of J1120+0641 based on the H$\alpha$ Balmer line, $M_{\text{BH}} = 1.52 \pm 0.17 \cdot 10^9 M_\odot$, which is in good agreement with the previous rest-UV Mg~{\small{II}} 
black hole mass measurement. 
The black hole mass based on the Paschen-series lines is also consistent within uncertainties, indicating that no significant extinction is biasing the $M_\text{BH}$ measurement obtained from the rest-frame UV. 
By comparing the ratios of the H$\alpha$, Pa-$\alpha$ and Pa-$\beta$ emission lines to predictions from a simple one-phase \textsc{Cloudy} model, we find that the hydrogen broad lines are consistent with originating in a common broad-line region (BLR) with density $\text{log} N_H/\text{cm}^{-3} \geq 12$, ionisation parameter $-7<\text{log} \ U<-4$, and extinction E(B--V)$\lesssim0.1$ mag. These BLR parameters are fully consistent with similarly-bright quasars at $0<z<4$.
Overall, we find that both J1120+0641's hot dust torus and hydrogen BLR properties show no significant peculiarity when compared to luminous quasars down to $z=0$. The quasar accretion structures must have therefore assembled very quickly, as they appear fully `mature' less than $760$ million years after the Big Bang.

\end{abstract} 
 

\section{Introduction}

The origin of the first supermassive black holes is an enduring puzzle. Measurements of the masses of supermassive black holes (SMBHs) based on their their rest-frame UV spectra have revealed the presence of a population of black holes with masses up to $M=1.5 \cdot 10^9$ M${}_\odot$ at $z>7$, less than $800$ million years after the Big Bang \citep{Mortlock11,Wang18,Yang19,Yang20,Yang21,Wang21}. The existence of such massive SMBHs requires either very massive black hole seeds of $\gtrsim 10^4$ M${}_\odot$ at $z=30$, or continuous growth rates in excess of the Eddington limit by a factor of a few (\citealt{Volonteri12,Pacucci17}; see \citealt{Inayoshi20} for a review). Both scenarios pose significant problems for current models of SMBH origins. The first accreting SMBHs, i.e.~quasars, have therefore been the target of extensive observational campaigns looking for clues to their growth mechanisms.

\begin{figure*}
\includegraphics[width=\textwidth]{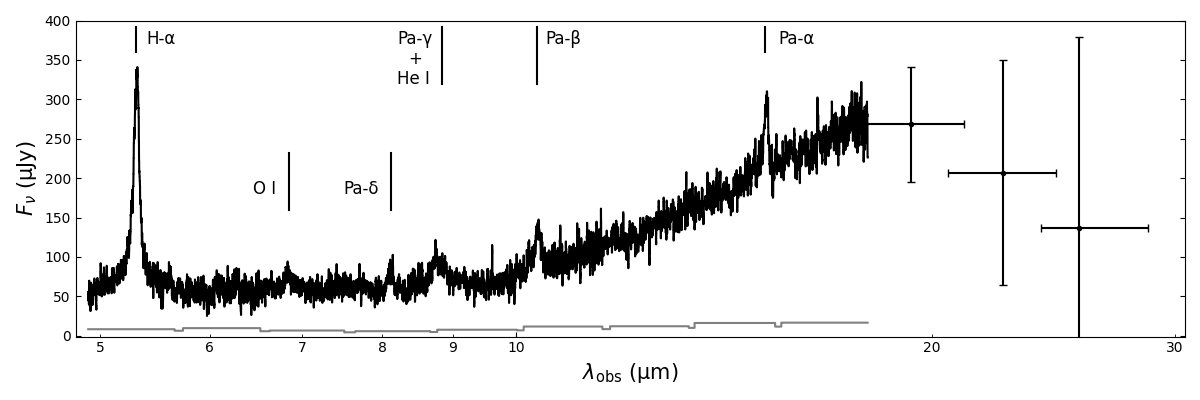}
\caption{MRS spectrum of J1120+0641 at $z=7.0848$. The detected emission lines are labelled.  The emission in Channel $4$, at $\lambda>17\mu$m, has been summed over the entire width of each sub-array due to the significantly lower sensitivity. The grey line indicates the $1\sigma$ uncertainty, which is calculated post-hoc on each individual subchannel (see text).}
\label{fig:spec}
\end{figure*}

Observations of $z>7$ quasars with optical and near-infrared instruments (corresponding to the quasars' rest-frame UV) have revealed a complex picture. 
On one hand, early quasars appear to be accreting matter at below or only slightly above the Eddington rate \citep{Shen19,Yang21,Farina22}.  
While the average Eddington ratios of early quasars appear to be higher than quasars with the same masses at $z<3$, the difference is due in large part to the selection bias towards brighter quasars at earlier times \citep{Farina22}. The rest-UV broad emission lines of  quasars at $5.7<z<7.5$ are identical in their luminosities and flux ratios to $2<z<5$ quasars, indicating unchanging levels of metal enrichment of quasar broad line regions (BLRs) across $2.6$ billion years of cosmic time \citep{Lai22}. 

On the other hand, the properties of quasar outflows do show signs of evolution at fixed quasar luminosity and black hole mass. The highly-ionised C~{\small{IV}} emission line displays outflows of $\lesssim1000$ km s$^{-1}$ on average in $1<z<5.5$ quasars, which more than double to $\sim2000-2500$ km s$^{-1}$ on average at $z>6.5$ \citep{DeRosa11,Mazzucchelli17,Shen19,Meyer19-qso,Schindler20}. Potentially related to the changes in the highly-ionised outflows, the occurrence rate of quasars with broad absorption line features (BAL quasars) among the bright quasar population more than doubles at $z>5.8$ compared to $2<z<4$ \citep{Yang21,Bischetti22,Bischetti23}.


These indications of changing quasar outflows have not yet yielded much progress in understanding early SMBH growth mechanisms. If SMBHs grow from massive initial seeds via accretion processes identical to late-time quasars, then presumably no changes at all should be seen in the earliest quasars. The main alternative to massive seeds is super-Eddington accretion in the earliest SMBHs, but changes in outflows are not a prediction of such models. In addition, no $z>6$ quasars are observed to be sufficiently super-Eddington. To circumvent this issue, models have often invoked significantly obscured (episodic) super-Eddington growth (e.g.
\citealt{Vito18,Trebitsch19,Davies19}). One advantage of episodic obscuration models is that they reconcile the short apparent lifetimes of $z>6$ quasars, some of which appear to be observed less than $0.1$ million years after the onset of the ongoing luminous phase \citep{Eilers18, Eilers20, Eilers21}. Such short obscuration-shifting timescales might (hopefully) leave imprints in the rest-infrared properties of the first quasars. In particular, remnants of obscuration could be visible (1) in the hot dusty torus, which is most tightly linked to accretion processes; (2) in obscuration of quasar BLR lines, which, if present, would affect the rest-UV $M_\text{BH}$ measurements employed to assess Eddington rates; (3) in the density and ionisation state of the BLR, as the `mode of accretion' shifts quickly from super-Eddington to sub-Eddington. The rest-frame infrared is therefore an ideal ground to search for signs of unusual quasar growth. We briefly elaborate on these 3 points below.

Quasars have long been known to display a prominent bump in emission around $1\lesssim \lambda_{\text{rest}} \lesssim 20 \mu$m, which has been associated with emission from dust grains \citep{Barvainis87,Kishimoto07,Mor12}. Based on reverberation mapping studies, the size of the dusty structure is known to be sub-parsec and to scale with luminosity \citep{Minezaki04,Suganuma06}, leading to the picture of a ``hot torus'' located near the dust sublimation radius, with temperatures $T>1000$K. The dusty torus has been modelled using both smooth dust distributions (e.g.~\citealt{Efstathiou95,Manske98,Schartmann05}) as well as more complex clumpy dust models (e.g.~\citealt{Schartmann08,Honig10}). Despite the potential complexity of the hot torus structure, observations have consistently found that the hot dust bump is adequately fit with a single-temperature black-body form across all quasar types and luminosities (e.g.~\citealt{Riffel09,Mor11,Landt13}). Some suggestions have made that quasars at $z\gtrsim6$ are more often hot-dust-free than their later-time counterparts \citep{Jiang10}, but these tentative results have not been confirmed \citep{Leipski14}. By pushing the observations of the hot torus dust to $z>7$, we will be able to establish whether the dust-enrichment of quasar accretion structures mimics the observed rapid metal enrichment.

The hydrogen Paschen-series emission lines are another crucial feature in the rest-frame infrared spectra of quasars \citep{Soifer04}. The Pa-$\alpha$ and Pa-$\beta$ lines are far less affected by extinction than hydrogen lines at shorter wavelengths, which makes more reliable tracers of black hole mass in highly extincted quasars \citep{Kim10}. 
The hydrogen Balmer and Paschen lines, in contrast with other broad lines in the rest-frame UV, optical and infrared, are non-resonant and give access to a common BLR via multiple transitions. 
As a consequence, the flux ratios of the lines are good tracers of potential dust extinction in the BLR, as well the strength of ionisation and density of the hydrogen BLR \citep{Kim10,Oyabu09}.

Both of these crucial observables, the hot torus dust and the hydrogen Paschen, are now available for the first time in a $z>7$ quasar with the JWST/MRS. 
The structure of the paper is as follows. We describe our MIRI-MRS observations of J1120+0641 and the data reduction procedure in Section~\ref{sec:obs}. Section~\ref{sec:methods} presents the methods employed in modeling the quasar's continuum emission (\S \ref{sub:meth_cnt}) and emission lines (\S \ref{sub:meth_lines}), as well our \textsc{Cloudy} model to interpret the emission line ratios (\S \ref{sub:meth_cloudy}). We present the results of our measurements in a similar order in Section~\ref{sec:results}. Section~\ref{sec:disc} contains a high-level summary of our results concerning the hydrogen BLR structure (\S \ref{sub:BLR}) and a discussion of the possibility of hot torus dust modeling beyond a black-body (\S \ref{sub:dust}). We summarise our results and conclude in Section~\ref{sec:ccl}. 

We employ a Planck cosmology throughout the paper \citep{Planck20}. All quoted rest-frame wavelengths correspond to their values in a \mbox{vacuum}.

\section{Observations}\label{sec:obs}

\subsection{The target}

The bright quasar J1120+0641 was the first to be discovered at $z>7$ \citep{Mortlock11}, and remained the only quasar known at $z>7$ for seven years. It was therefore the target of extensive multi-wavelength observations which justify its inclusion as a priority target for GTO Cycle 1 observations. At the time of writing, nine quasars are now known at $z>7$ (e.g.~\citealt{Banados18,Matsuoka19,Wang21}; see \citealt{Bosman20-list} for an up-to-date list). 

Observations of J1120+0641 with NOEMA \citep{PdBI} and later ALMA \citep{ALMA} have detected sub-mm continuum and [C~{\small{II}}] fine-structure line emission from its host galaxy at a systemic redshift of $z=7.0848 \pm 0.0004$ \citep{Venemans12, Venemans20}. A deep spectrum of J1120+0641 was obtained with the X-Shooter instrument \citep{XSHOOTER} in the optical and at NIR wavelengths \citep{Bosman17}. It reveals a fast outflowing C~{\small{IV}} emission line with a velocity  $v=2136^{+32}_{-27}$ km s$^{-1}$
with respect to systemic \citep{Schindler20}, a value which is significantly elevated compared to $z<5$ quasars of comparable brightness \citep{Bosman15}, but average among $z>6$ quasars. 

The black hole mass of J1120+0641 is $1.35\pm0.04 \cdot 10^9 M_\odot$ based on the Mg~{\small{II}} line (\citealt{Yang21}; see Section~\ref{sub:res_hlines} for more details), and its Eddington ratio is $\sim 0.40$ \citep{Farina22}. J1120+0641 is radio-quiet \citep{Momjian14,Banados15} and its X-ray properties are indistinguishable from other bright quasars at $1<z<5$ with similar Eddington rates \citep{Page14,Moretti14,Wang20}.

\subsection{MRS Observations}

JWST \citep{Gardner+23} observations of J1120+0641 were obtained in January 12, 2023 (proposal ID 1263, P.I.~Luis Colina) using the Mid-Infrared Resolution Spectrograph (MRS, \citealt{Wells+15, Argyriou+23}) of the Mid-Infrared Instrument (MIRI, \citealt{Rieke+15, Wright+15, Wright+23}). The spectrograph provides integral field spectroscopy covering the 4.9 to 27.9 $\mu$m range with intermediate spectral resolution \citep{Wells+15, Labiano+21, Jones+23}. The observations using the three MRS configurations (SHORT, MEDIUM, and LONG) covered the entire MRS spectral range. The  total integration time of 3153 seconds per configuration was divided into 12 integrations, each using the SLOWR1 readout mode with 16 groups. A 4-point dither pattern designed for point-sources and three integrations per dither were used. 

\subsection{Reduction process}

The MRS observations were processed with version 1.10.2 of the JWST calibration pipeline, and version 11.17.0 and context 1089 of the Calibration Reference Data System (CRDS). In general, we follow the standard MRS pipeline procedure (\citealt{Labiano+16}; and \citealt{Alvarez-Marquez+23a, Alvarez-Marquez+23b} for examples of MRS data calibration in low and high redshift galaxies). In addition, we have included customized steps to improve the quality of the final MRS calibrated products.

The first stage of the MRS pipeline (\citealt{Argyriou+23}; Morrison et al. in prep.), which performs the detector-level corrections and transforms the ramps to slope detector images, was run making especial attention to identify and correct the cosmic ray (CR) showers. We have turned on the \texttt{find\_showers} keyword of the \texttt{jump} pipeline step, and use 60 seconds and 400 pixels for the \texttt{time\_masked\_after\_shower} and \texttt{max\_extended\_radius} keywords. It identifies and corrects the majority of faint CR showers, but the brightest CR showers, which generate a latency in the MIRI detectors, are still present in the final slope images. We identify warm pixels on the detector based on their variation in time and include them in the bad pixel mask. We have derived the median of all slope images of each detector, independent of the pointing and MRS band. In each median detector image, we have applied a classical sigma clipping algorithm to identify the warm and bad pixels. The detected warm and bad pixels that were not previously identified by the bad pixel mask, were updated as \texttt{do\_not\_use} pixels in the data quality map of each individual slope image.    

The second stage of the pipeline, which implements the specific calibrations for the MRS observing mode \citep{Argyriou+23,Patapis+23,Gasman+23}, was run as default but using new flat-field, fringe-flats, and \texttt{photom} reference files derived from the \textit{JWST} Cycle 1 calibration programs. These reference files are publicly available in CRDS. We have performed a pixel-by-pixel background subtraction after stage 2 of the pipeline. Our MRS observations were performed without dedicated background observations, therefore we take advantage of the dither strategy to generate a master background using the on-source observations. In each detector calibrated image, we have identified the slices that are affected by the emission of the source, and they have been masked. Then, we have generated the master background image of each MRS band by performing a median of all available exposures. The master background images of each band have been subtracted from the original calibrated detector images.

The third stage of the pipeline, which combines the different exposures within an observation to generate a final MRS 3D spectral cube \citep{Law+23}, was run using the background subtracted and fully calibrated detector images and skipping the \texttt{master background subtraction} and the \texttt{outlier detection} steps. This process produced twelve 3D spectral cubes, one for each band (SHORT, MEDIUM, and LONG) of the MRS channels, with spatial and spectral sampling of 0.13"\,$\times$\,0.13"\,$\times$\,0.8\,nm for channel one, 0.17"\,$\times$\,0.17"\,$\times$\,1.3\,nm for channel two, 0.20"\,$\times$\,0.20"\,$\times$\,2.5\,nm for channel three, and 0.35"\,$\times$\,0.35"\,$\times$\,6\,nm for channel four. The resolving power of the MRS observations ranges from 4000, in channel one, to 1500, in channel four, corresponding to a FWHM of 75 km\,s$^{-1}$ to 200 km\,s$^{-1}$ respectively \citep{Labiano+21, Jones+23}. Third-level (pixel-level) defringing was applied, but its impact was minimal.

We have performed 1D spectral extractions individually in each of the MRS cubes using a circular aperture of radius equal to $1 \times$ FWHM $(\lambda)$, where FWHM $(\lambda) = 0.3$ arcsec for $\lambda < 8\mu$m and FWHM $(\lambda) = 0.31\times \lambda[\mu m] / 8$ arcsec for $\lambda > 8\mu$m. The selected FWHM\,($\lambda$) values follow the MRS Point Spread Function (PSF) Full Width at Half Maximum (FWHM). We use the MRS PSF models (Patapis et al.~in prep.) to correct the aperture losses in the 1D extracted spectra. The percentage of flux that is lost out of the selected aperture is $\sim\,$30\% for all MRS channels. Finally, the channels were stitched together by interpolating the flux over the overlapping wavelength regions and averaging without weighing.



\begin{figure*}
\includegraphics[width=\textwidth]{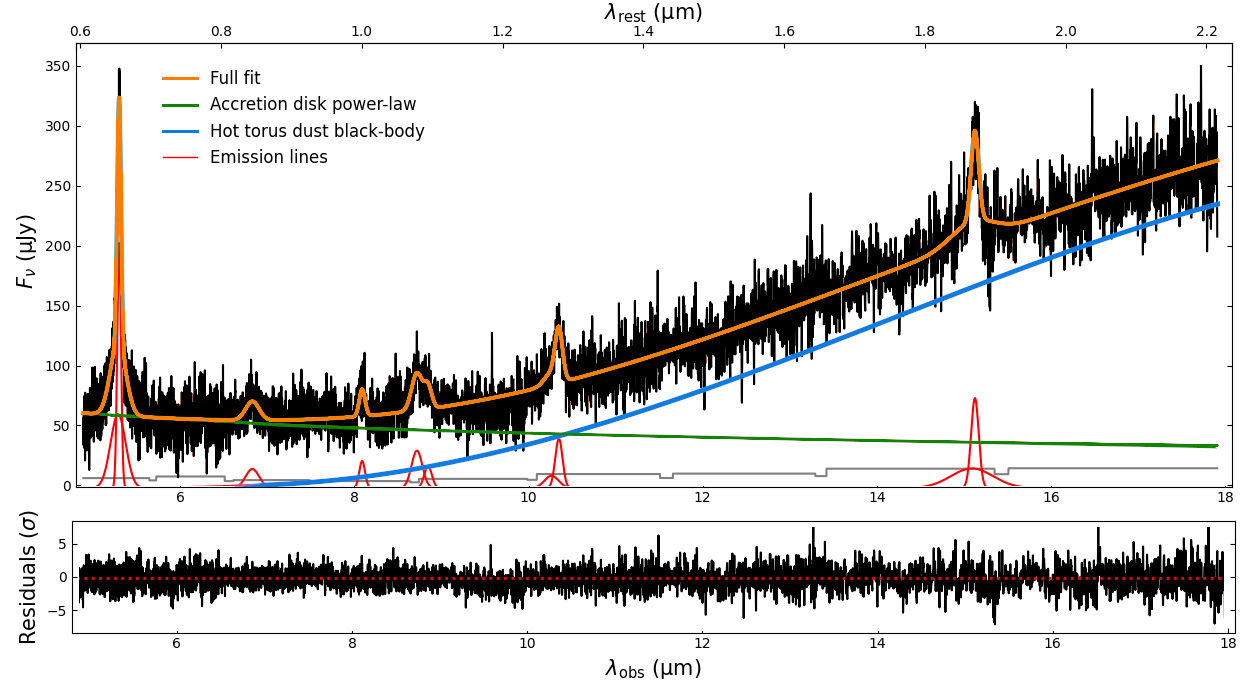}
\caption{Full fit to the MIRI-MRS spectrum of J1120+0641 (top, orange) and its residuals (bottom) in units of standard deviations. The coloured lines show the emission coming from the accretion disk (power law, blue), the hot torus dust (black-body radiation, green) and the broad quasar emission lines (red). The residuals show no signs of deviation from the continuum model, nor additional broad emission lines. The gray line is an in Fig.~\ref{fig:spec} and the red dashed line indicates the location of zero residuals. See Fig.~\ref{fig:spec} for line IDs.}
\label{fig:fit}
\end{figure*}

\subsection{Uncertainties}

Due to the complexity of the reduction process, it was not possible to obtain the spectroscopic uncertainties from first principles concurrently with the spectral extraction. Instead, we estimate the uncertainties \textit{a posteriori} using the following empirical procedure. After a first pass of the fitting procedures (see \textit{Methods}, below), we make the assumption that the fitting residuals are well described by a random gaussian distribution. For each MRS subchannel, we fit a gaussian distribution to the residuals and use its full width at half maximum (FWHM) as an estimate of the uncertainty over the whole subchannel. The spectral uncertainties are averaged and divided by $\sqrt{2}$ over the overlapping edges of the subchannels to reflect the improved SNR in those regions. 

We tested the effect of varying the empirically-estimated uncertainties on our fitted parameters by arbitrarily rescaling the uncertainties by factors up to $3$, finding no statistically significant effect on the recovered optimal parameter values. The parameter uncertainties, however, are affected by this rescaling. However, since the final fit residuals do appear indistinguishable from a gaussian distribution, we consider our empirical uncertainty estimation method to be sufficient for our purposes. The need to model uncertainties post-hoc will probably disappear with future improvements to the official pipeline.


\section{Methods}\label{sec:methods}

We aim to derive the black hole mass, the dust temperature in the torus and the physical properties of the BLR of J1120+0641. To achieve this we need a model-fitting method flexible enough to fit the continuum and the complex emission line profiles simultaneously, and derive covariant uncertainties between all parameters. We therefore employ the \textsc{Sculptor} software, which provides an open-source framework for the viewing and analysis of quasar spectroscopy \citep{sculptor}.

We added custom functions to \textsc{Sculptor} for the modeling of the rest-infrared continuum emission, and used built-in functions for the fitting of the emission lines. \textsc{Sculptor} provides a convenient way of specifying bounds for all model parameters and for tying parameters together, as described in the following sub-sections. We use the built-in Monte-Carlo Markov Chain (MCMC) sampler with $200$ walkers and $50$ rounds of burn-in to generate parameter constraints and covariance matrices.  
All model parameters are fit concurrently. 


\subsection{Continuum emission}\label{sub:meth_cnt}

Quasar continuum emission at $0.09 \lesssim \lambda_{\text{rest}} \lesssim 1 \mu$m is dominated by radiation from the accretion disk, which is empirically well described by a power-law
\begin{equation}
F_\lambda \propto \lambda^\beta. 
\end{equation}
Over this wavelength range, multiple spectral breaks are known to be present, most prominently at $\lambda_{\textit{rest}} \sim 0.12 \mu$m, $\lambda_{\textit{rest}} \sim 0.3 \mu$m and $\lambda_{\textit{rest}} \sim 0.6 \mu$m \citep{Glikman06,Hernan16}. Since our MRS spectrum covers the range $\lambda_{\textit{rest}}>0.61 \mu$m, a single unbroken power-law is a sufficient description of the accretion disk emission. We set a broad flat prior on the power-law slope of $\beta \in [-2.5,2.5]$, and no prior on the power-law amplitude. The power-law parameters are fit together with all other model parameters over the entire wavelength range except for minor masking of fringing features (see later).

Based on observations of a strong emission bump in the rest-IR, hot dust is known to be present in the accretion structure surrounding the accretion disk, a.~k.~a.~the torus. 
The torus dust is traditionally assumed to be highly optically thick, and therefore it is usually modeled as a pure black body: 
\begin{equation}
F_\nu \propto \frac{M_{\text{dust}}}{M_\odot} \lambda^{-3} \left(\exp\left(\frac{h c}{k \lambda T}\right) -1\right)^{-1}. 
\label{eq}
\end{equation}
Hot torus dust is generally thought to consist of graphitic dust due to the lack of strong dust absorption features seen in obscured quasars and the fact that silicate dust sublimates at $T\sim1300-1500$K \citep{Salpeter77,Lodders03} while hot dust temperatures of $T>1700$K are seen in Seyfert galaxies (e.g.~\citealt{Landt15}). Graphitic hot torus dust instead sublimates at temperatures $T\sim2000$K, meaning that it can start to dominate the quasar spectrum at $\lambda \gtrsim 0.9 \mu$m. Meanwhile, black-body-emitting dust colder than $T\sim300$K will not emit significantly in the range covered by MIRI-MRS Channels $1$, $2$ or $3$ unless the (surface) mass of dust is unphysically large ($M_{\text{dust}} > 10^5 M_\odot$). Consequently, we set a prior on $T \in [300,3000]$ K.

\begin{figure*}
\includegraphics[width=\textwidth]{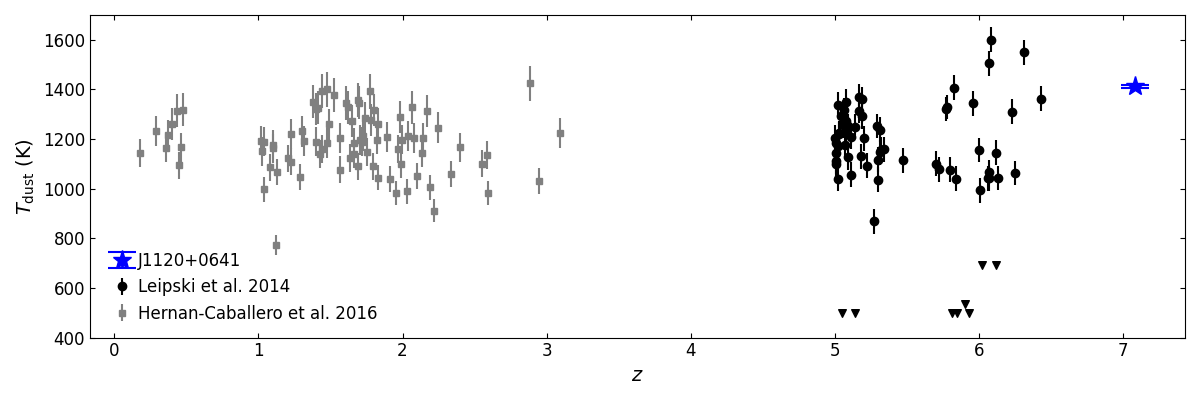}
\caption{Measurements of hot torus dust temperature across redshift. Uncertainties of $\pm5\%$ are assumed for the measurements of \citet{Hernan16}. 
Black triangles indicate upper limits from \citet{Leipski14}. Our spectroscopic dust temperature measurement in J1120+0641 has significantly smaller uncertainties. The occurrence rate of quasars with $T_{\text{dust}} > 1400$K might be increasing towards higher redshifts, as $4/16$ quasars at $z>6$ have dust than $1400$K compared to $1/137$ at $z<6$.}
\label{fig:evo}
\end{figure*}

Under the assumption of pure black-body radiation, the mass of dust cannot be constrained without a model for its physical distribution, since only the radiating surface of the dust torus is assumed to be visible. The factor $M_{\text{dust}}$ in Equation~\ref{eq} is therefore to be better understood as a dust surface area, but we retain the notation for consistency with older works and with the optically-thin regime (see below). Because the dependence of Equation~\ref{eq} on $T$ is exponentially stronger than on $M_{\text{dust}}$, we set a logarithmically-flat prior on $\log \left(M_{\text{dust}}/M_\odot \right) \in [-5,5]$. 


\subsubsection{More complicated dust models}

It is not known \textit{a priori} whether hot torus dust emission is indeed optically thick. In principle, an extra modified black-body term $\left(1 - \exp(-\tau(\nu))\right)$ can be added as a prefactor to Equation \ref{eq}, where $\tau(\nu)$ is a frequency-dependent optical depth $\propto \nu$. Past studies were generally limited in fitting spectral energy distributions (SEDs) to a small number of photometric points, and both the optically-thin and optically-thick SED forms have been employed in the literature to constrain the temperature. The optically-thick formula shifts the location of the emission's peak to longer wavelengths, resulting in larger temperature estimates than the optically-thin assumption. 
The significantly higher resolution and sensitivity of MIRI-MRS compared to past FIR instruments could in principle distinguish between the optically thick and thin solutions using the shape of the continuum upturn, motivating us to conduct a separate fitting attempt using the optically-thin formula. 

The proportionality factor between $F_\nu$ and $M_{\text{dust}}$ then depends on the physical cross-section of the emitting object, which is unknown for a quasar hot dust torus. The dust mass therefore still cannot be constrained unless we make the assumption of fully optically thin dust ($\tau \to 0$), in which case the spectrum becomes:
\begin{equation}
F_\nu \propto \frac{M_{\text{dust}}}{M_\odot} \lambda^{-(3+\beta)} \left(\exp\left(\frac{h c}{k \lambda T}\right) -1\right)^{-1}, 
\label{eq2}
\end{equation}
where $\beta$ is the modified black-body slope, usually taken to be between $1.5\leq\beta\leq2.0$. We conduct fits to the data using both a fixed $\beta=2$ and a free $\beta \in [1.5,2]$.

The exponential dependence of the emission on temperature causes the rest-IR spectrum to be dominated by a relatively much smaller amount of the hottest dust which is present. We therefore begin by considering dust of a single temperature, corresponding to the hottest dust present at the inner edge of the accretion torus. This single-temperature assumption is natural in the black-body radiation case, since only radiation originating from the surface of a very dense torus is assumed to be visible. However, interferometric observations of low-$z$ AGN, e.g.~with GRAVITY \citep{GRAVITY}, have revealed that (somewhat colder) torus dust can be significantly entrained along the polar axis \citep{Lopez-Gonzaga16,Konig17}. The presence of a small amount of optically-thin hot dust is therefore plausible in principle for luminous quasars (but see \citealt{Dexter20,Pfuhl20} for the lack of such an effect in fainter AGN, and see also results from the Matisse instrument revealing more complicated dust distribution in \citealt{Rosas22}).

However, since we find no statistically significant deviation of the spectrum from the one-temperature continuum (see Section~\ref{sub:res_cnt}), we do not consider more complicated dust models where the emission arises from a distribution of dust masses and temperatures. Such models would entail a significant increase in the number of model parameters which we would not be able to meaningfully constrain, and their interpretation would be anyway hindered by our lack of knowledge of $\tau(\nu)$. 


\subsection{Emission lines}\label{sub:meth_lines}

Broad quasar emission lines from the BLR are known to contain multiple gaussian (or lorenzian) components of different widths, with the broadest components typically showing the strongest outflows with respect to systemic redshift. Accordingly, we try to model the broad lines with two gaussian components when the data quality allows it. More details on individual broad lines are given below. The broad line parameters were fit together with the continuum emission parameters in order to take potential covariances into account in the posteriors. 

We calculate emission line fluxes $f_\text{X}$ by integrating the (double) gaussian best-fit models, and the line luminosities via $L_\text{X} = 4 \pi D_L^2 f_\text{X}$. The bolometric luminosity of J1120+0641 is $L_\text{bol} = 1.705 \pm 0.005 \cdot 10^{47} \text{erg s}^{-1}$, as calculated by \citet{Farina22} based on the quasar's luminosity at $3000$\AA \ using the empirical relation of \citet{Richards06}, updated by \citet{Shen11}:
\begin{equation}
L_\text{bol} = 5.15 \times \left( 3000 \text{\AA} \ L_{\lambda, 3000 \text{\AA}} \right) .
\end{equation}
Finally, the Eddington ratio of accretion in J1120+0641 is obtained via 
\begin{equation}
\lambda_\text{Edd} = \frac{L_\text{bol}}{L_\text{Edd}} = \frac{L_\text{bol}}{1.257\cdot10^{38}} \left( \frac{M_\text{BH}}{M_\odot} \right)^{-1} ,
\end{equation}
where $L_\text{Edd}$ is the Eddington accretion rate for a black hole with mass $M_\text{BH}$. 
The intrinsic scatter in the $L_\text{bol}$ scaling relation is a factor of $\sim2$, dominating the error budget compared to measurement uncertainties. However, by using the same value of $L_\text{bol}$ as past literature, we ensure that comparisons of $\lambda_\text{Edd}$ across multiple emission lines are meaningful.

\begin{table*}
\begin{center}
\caption{Best-fit parameters for the broad emission lines detected in J1120+0641's MIRI-MRS spectrum (in the rest-frame of the quasar). The H$\alpha$, Pa-$\alpha$ and Pa-$\beta$ lines are fitted with two gaussian components, labeled BC1 and BC2 for each transition. All parameters are simultaneously free to vary within the priors given in the text, except for the centre of the Pa-$\alpha$ BC2, which is tied to the same redshift as H$\alpha$'s BC2 due to the presence of nearby fringing residuals which hamper the fit.} 
\begin{tabular}{l | l l l c c c}
Line & Peak  & $z$ & FWHM  & $f$ & $L$ & Notes \\
 & ($\mu$m) & & (km s$^{-1}$) & ($10^{-16}$ erg s$^{-1}$ cm$^{-2}$) & ($10^{44}$ erg s$^{-1}$) &  \\
\hline
\hline
H$\alpha$ total & $5.3124_{-0.0002}^{+0.0002}$ & $7.095 \pm 0.002$ & $3430 \pm 180$ & $26.50 \pm 0.30$ & $16.23\pm0.18$ & \\
H$\alpha$ BC1 & $5.3129_{-0.0006}^{+0.0007}$ & $7.0933\pm 0.0010$ & $2880_{-110}^{+140}$ & & &\\
H$\alpha$ BC2 & $5.3009_{-0.0014}^{+0.0016}$ & $7.075\pm 0.002$ & $11720_{-710}^{+560}$ & & &  \\
\hline
Pa-$\alpha$ total & $15.183_{-0.004}^{+0.008}$ & $7.095_{-0.002}^{+0.004}$ & $2220^{+290}_{-190}$ &  $2.81 \pm 0.26$& $ 1.88 \pm 0.17$ &   \\
Pa-$\alpha$ BC1 & $15.185 \pm 0.006$ & $7.096 \pm 0.004$  & $2000_{-260}^{+190}$ & & & \\
Pa-$\alpha$ BC2 & \textit{15.1365} & \textit{7.075} & $11300_{-2800}^{+2000}$ &  & & Centroid held fixed \\
\hline
Pa-$\beta$ total & $10.384\pm0.004$ & $7.098\pm 0.003$ & $2960^{+190}_{-100}$ & $1.82\pm 0.37$ & $1.11\pm0.22$ &   \\
Pa-$\beta$ BC1 & $10.385 \pm 0.003$ & $7.099\pm 0.002$ & $2830_{-130}^{+190}$ & & &  \\
Pa-$\beta$ BC2 & $10.312 \pm 0.005$ & $7.042\pm 0.004$ & $10340_{-2500}^{+1800}$ &  & & Affected by fringing\\
\hline
Pa-$\gamma$ & $8.874_{-0.004}^{+0.003}$ & $7.111 \pm 0.004$ & $3520_{-390}^{+440}$ &$0.94\pm0.35$ & $0.57\pm0.21$&  Blended with He~{\small{I}} \\
Pa-$\delta$ & $8.1156_{-0.0015}^{+0.0025}$ & $7.0736_{-0.0015}^{+0.0025}$ & $2880 \pm 100$ & $ 0.87\pm0.31$ & $ 0.53\pm0.19$ &  \\
He~{\small{I}} & $8.7477_{-0.0029}^{+0.0022}$ & $7.077^{+0.002}_{-0.003}$ & $4370_{-350}^{+410}$ &   $1.85\pm0.29$ & $ 1.14\pm 0.18$& Blended with Pa-$\gamma$ \\
O~{\small{I}} & $6.849_{-0.003}^{+0.003}$ & $7.109\pm0.003$ & $8000_{-1100}^{+800}$ & $2.05 \pm 0.14$& $ 1.25 \pm 0.10$  &  \\
\end{tabular}
\label{tab:res}
\end{center}
\end{table*}

\subsubsection{H$\alpha$}

The broad H$\alpha$ line is modeled with two gaussian components with free centroids, amplitudes, and widths, amounting to $6$ free parameters. We initiate the fit with a reasonable first guess obtained by-eye in the \textsc{Sculptor} interface. Both broad gaussian components are given broad flat priors of centroid $\in [5.2, 5.4] \mu$m, width $\in [300, 20000]$ km s$^{-1}$ and unconstrained amplitude. 
We find that all parameters are well constrained with no need for the presence of an additional narrow-line component of H$\alpha$, and no signs of [N~{\small{II}}] 
emission. The lack of stellar emission lines is consistent with the relative faintness of the host galaxy in sub-mm [CII] \citep{Venemans20}. The host galaxy has an estimated mass of $\sim4.3 \cdot 10^{10} M_\odot$, only twenty times more massive than the SMBH \citep{Venemans17}. We therefore do not include an additional narrow component for the H$\alpha$ line. 


Black hole masses based on the broad H-$\beta$ line are well-calibrated from reverberation-mapping studies (e.g.~\citealt{Shen23}). Since the widths of the H$\alpha$ and H-$\beta$ lines follow a tight relationship, H$\alpha$ may also be used to calculate the black hole virial mass on its own \citep{Shen11, Greene05}:
\begin{equation}
\begin{split}
\log \left( \frac{M_{\text{BH, H-}\alpha}}{M_\odot} \right) = 0.379 +0.43 \log \left( \frac{L_{\text{H-}\alpha}}{10^{42} \text{erg s}^{-1}} \right) \\
+2.1 \log \left( \frac{\text{FWHM}_{\text{H-}\alpha}}{\text{km s}^{-1}} \right) .
\end{split}
\label{eq:ha}
\end{equation}
This scaling relation possesses an intrinsic systematic scatter of a factor of $\sim2$, which is therefore the dominating uncertainty in our measurement, far above the uncertainties in the parameters $L_{\text{H-}\alpha}$ and FWHM$_{\text{H-}\alpha}$ which are calculated from emission line fitting. The FWHM$_{\text{H-}\alpha}$ is the width of the total line model where its value is equal to half of its maximum.


\subsubsection{Pa-$\alpha$} 

Similarly to H$\alpha$, we attempt to model the Paschen-$\alpha$ $1.8756 \mu$m line with two broad gaussian components. While the central component is well-constrained, the broad component's centroid does not converge. We suspect that this is caused by low-frequency fringing in the spectrum, which is visible at $15.5<\lambda<16.5 \mu$m and may extend over the neighbouring Pa-$\alpha$ emission line (see Figure \ref{fig:pa}). In order to surmount this difficulty, we fix the redshift of the broadest line component to be the same as for the H$\alpha$ emission line. The centroid of the other component is given a flat prior $\in [15.0,15.5] \mu$m, with the same width and amplitude priors as the H$\alpha$ emission line. 

A black hole mass estimate can be obtained from the Pa-$\alpha$ directly \citep{Kim10}:
\begin{equation}
    \frac{M_{\text{BH, Pa-}\alpha}}{M_\odot} = 10^{7.31} \left(\frac{L_{\text{Pa-}\alpha}}{10^{42} \text{erg s}^{-1}}\right)^{0.48} \left(\frac{\text{FWHM}_{\text{Pa-}\alpha}}{10^3 \text{km s}^{-1}}\right)^{1.68}
\end{equation}
where $L_{\text{Pa-}\alpha}$ and FWHM$_{\text{Pa-}\alpha}$ are the luminosity and the full width at half maximum of the line, respectively. The intrinsic scatter of the Pa-$\alpha$-based black hole mass estimate is potentially very low ($\sim0.2$ dex, \citealt{Kim10}), but it has been tested on fewer black holes with known mass from reverberation mapping compared to scalings based on H$\alpha$.

\subsubsection{Pa-$\beta$}

The Paschen-$\beta$ $1.2822 \mu$m line is similarly modelled with two gaussian components, with the same amplitude and width constraints as the previously discussed lines, and a prior on the gaussian centroids $\in [10.3,10.5]\mu$m. The presence of a spurious fringing feature (or unidentified narrow emission line) at $\lambda = 10.223 \mu$m biases the fitting, and therefore we mask this feature. The fit is shown in Figure.~\ref{fig:pb}. 

A recent black hole mass estimator based on the Pa-$\beta$ line is given by \citet{Kim15}:
\begin{equation}
    \frac{M_{\text{BH, Pa-}\beta}}{M_\odot} = 10^{7.04} \left(\frac{L_{\text{Pa-}\beta}}{10^{42} \text{erg s}^{-1}}\right)^{0.48} \left(\frac{\text{FWHM}_{\text{Pa-}\beta}}{10^3 \text{km s}^{-1}}\right)^2.
\end{equation}
Similarly to the estimate based on the Pa-$\alpha$ line, the Pa-$\beta$ estimator has a low intrinsic scatter of $\sim0.2$ dex, but has been tested on smaller samples of quasars possessing reverberation mapping measurements of the black hole mass \citep{Kim10}.

\subsubsection{Other emission lines}

We model the Paschen-$\gamma$ $1.0941\mu$m $+$ He~{\small{I}} $1.0830\mu$m complex at $\lambda \sim 8.82 \mu$m with two gaussians (see Fig.~\ref{fig:pg}). In order to distinguish the two components, we constrain the centroid of the He~{\small{I}} component to be located $\in [8.6,8.77] \mu$m and the centroid of Pa-$\gamma$ to be $\in [8.77,9.0] \mu$m. Due to the relatively low SNR of $\sim 4.5$ per pixel, the fit posterior does not completely drop at the boundary between the centroids; i.e.~the components are not fully separated at the $3 \sigma$ level. However, we are still able to obtain posteriors on the line widths at $2 \sigma$. 

The remaining detected emission lines, O~{\small{I}} $0.8446\mu$m and Pa-$\delta$ $1.0052 \mu$m, are modelled with single gaussians and are well-constrained (Fig.~\ref{fig:oi}).

It is possible to derive a black hole mass directly from the rest-IR Paschen-series lines and the rest-IR continuum. Following the method of \citet{Landt13}, the Pa-$\alpha$ and Pa-$\beta$ lines can be calibrated to reverberation mapping measurements, which yields the relation:
\begin{equation}
\begin{split}
    \log M_{\text{BH, IR}} = \left(0.88 \pm 0.04\right) \cdot \bigg( 2 \log \frac{\text{FWHM}_{\text{H}}}{\text{km s}^{-1}} \\
    + 0.5 \log \frac{\nu L_{1\mu\text{m}}}{\text{erg s}^{-1}} \bigg) - \left( 17.39 \pm 1.02 \right) ,
    \end{split}
\label{eq:pa}
\end{equation}
where FWHM$_{H}$ is the FWHM of one/multiple Paschen-series hydrogen lines (under the assumption that they originate in a common BLR). Although this relation possesses $1$ dex of intrinsic scatter, we will be able to check its consistency with other BH mass tracers.

\subsection{Cloudy modeling}\label{sub:meth_cloudy}

In order to extract information on the hydrogen BLR from the ratio of the H$\alpha$ and Paschen-series lines, we implement a simple model of the BLR with the software {\textsc{Cloudy}} \citep{Cloudy}. We use a simplified version of the model proposed by \citet{Tsuzuki06}, where frequency-dependent emission $F(\nu)$ is incident on a single BLR cloud with density $N_H$. 
The form of $F(\nu)$ consists of a UV bump, power-law emission over the X-rays, and a high-energy cut-off:
\begin{equation}
\begin{split}
    F(\nu) \propto \nu^{-0.2} \rm{e}^{-h\nu/kT_{\text{cut}}} \rm{e}^{-0.136{\text{eV}}/h\nu} \\ 
    + \alpha \nu^{-1.8} ,
\end{split}
\end{equation}
where $T_{\text{cut}} = 150,000$K and $\alpha$ is normalised such that $F(2\text{keV})/F(2500\text{\AA}) = 2.25\cdot10^{-4}$. The values of the UV slope and X-ray cut-off are consistent with observations of J1120+0641 within uncertainties. We parametrise the strength of the incident emission with the parameter $U$ (the ``ionisation parameter'') such that $U=\Phi/cN_H$ where $\Phi (\text{s}^{-1} \text{cm}^{-2})$ is the photon flux. Emission line ratios remain nearly constant with $U$. We refer interested readers to \citet{Tsuzuki06} for more details of the model's physical derivation, and to \citet{Oyabu09} for an application to the hydrogen BLR. Unlike those authors, we do not include metals in the BLR cloud, and we assume no turbulence of the gas. 
We record the output of the model for three values of $\text{log} N_H (\text{cm}^{-3}) = 10, 12, 14$ and values of $\text{log} U = -7, -6, ..., 0$.


\section{Results}\label{sec:results}

Figure~\ref{fig:fit} shows the full fit to the spectrum over Channels $1$, $2$ and $3$, as well as the fit's residuals in units of standard deviations. We do not use the data from Channel 4, since the quasar is only marginally detected there. We find that the residuals are adequately described by a gaussian distribution over all Channels, with no signs of additional broad emission lines or slow-varying deviations. The best-fit parameters for the broad emission lines and their uncertainties are given in Table~\ref{tab:res}.

\subsection{Continuum emission and torus dust temperature}\label{sub:res_cnt}

\begin{figure}
\includegraphics[width=\columnwidth]{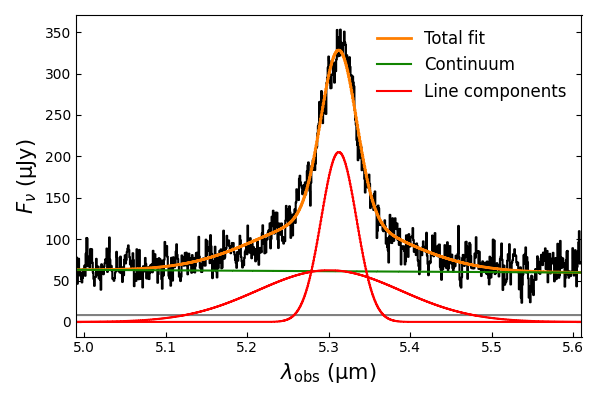}
\caption{Spectrum of the H$\alpha$ broad line (flux in black and uncertainty in grey) with the optimal line model fit (orange), split into the continuum constituting of a power-law and a black-body (green) and the broad emission lines (red).}
\label{fig:ha}
\end{figure}

The slope of the accretion disk continuum is fit as $\beta=-0.430_{-0.027}^{+0.022}$, indicating a typical ``blue'' continuum with no sign of extinction of the accretion disk. The emission from the accretion disk becomes equal to the brightness of the hot torus dust around $\lambda_{\text{obs}} = 10.5\mu$m (Figure~\ref{fig:fit}). A hot dust temperature of $T = 1413.5_{-7.4}^{+5.7}$K provides the best fit to the continuum under the assumption of black-body emission.


The black-body temperature is in relatively good agreement with measurements of torus dust in quasars at $5<z<6.5$ (Figure~\ref{fig:evo}). \citet{Leipski14} estimate the uncertainties in photometry-based measurements to be around $\pm 50$K; in contrast, the spectroscopic measurement enabled by the MIRI-MRS is accurate to better than $10$K ($0.5\%$). Unlike previous claims of the existence of hot-dust-free quasars at $z>5$, we do not detect a deficit of dust in J1120+0641. The temperature we measure is slightly elevated compared to past literature, with only $3$ quasars out of \citet{Leipski14}'s sample of $68$ having higher temperatures ($4\%$ of the sample). Adding J1120+0641 to $3$ quasars with $T>1400$K are exclusively found at $z>6$, where they constitute $25\%$ of all quasars ($4/16$). In contrast, no quasars with $T>1400$K are found at $z<6$ ($0/53$). This suggests a possible evolution in dust temperature. To estimate the statistical significance of the evolution, we use the Poisson occurrence statistics tables of \citet{Gehrels86}. The occurrence rate of quasars with $T>1400$K dust at $z<6$ is constrained to be $<0.071 (<0.124)$ at $2\sigma (3\sigma)$. Meanwhile, the occurrence rate above $z=6$ is $0.25^{+0.22}_{-0.18} \left({}^{+0.55}_{-0.22}\right)$ at $2\sigma (3\sigma)$. Based on these calculations, the evolution in quasar dust temperature is statistically significant at more than $2\sigma$ significance, but less than $3\sigma$. All past studies have similarly assumed black-body emission.

Measurements of the hot dust temperature in $84$ quasars at $z<5$ are provided by \citet{Hernan16} based on low-resolution spectroscopy ($R\sim20$) from \textit{Spitzer} and AKARI.
Based on the difference between the temperatures measured by \citet{Hernan16} and \citet{Leipski14} in the only quasar where the sample overlaps (J1148+5251 at $z=6.4$), we estimate uncertainties of about $5\%$, which are plotted in Fig.~\ref{fig:evo}. At $z<5$, only $1$ quasar out of $84$ shows a temperature $T_{\text{dust}}>1400$K. Combining the samples, the temperature trend becomes even stronger: $4$ out of $16$ quasars studied at $z>6$ have dust hotter than $1400$K, and only $1/137$ over $0<z<6$, for similar quasar bolometric luminosities. The significance of the trend, however, remains below $3\sigma$.




However, a potential caveat is the lack of correction for contamination by broad emission lines in past studies. The Paschen-$\alpha$ and Paschen-$\beta$ lines, in particular, contribute flux to the same observed wavelengths as the hot dust ``upturn" which is used for SED fitting the temperature in some past photometric studies (e.g.~\citealt{Leipski14}). The effect of contamination by emission lines would be a systematic under-estimation of the dust temperature, which may explain the shift between J1120+0641 and some of the past literature. We plan to quantify this bias and explore further trends of the dust temperature with, e.g.~black hole mass, in future work. More spectroscopic observations of hot dust in $z>6$ quasars, but also in $z<6$ quasars, are also clearly needed to confirm this potential temperature trend. 


The potential cause(s) of an evolution towards hotter tori at earlier times is unclear. J1120+0641's Eddington rate, luminosity, black hole mass, and X-ray properties are normal compared to the lower-redshift comparison samples. We speculate that a change in dust composition could create such an effect in principle. Both graphite and silicate grains have been suggested to make up quasar tori, since they can withstand the high measured temperatures without sublimating. Small dust grains could in principle lead to a higher torus density and therefore a higher temperature at the innermost radius, but small grains also sublimate more readily. Further modeling work is required to quantify the potential impact of dust make-up on measured torus temperature.


\subsection{Emission-line black hole mass estimates}\label{sub:res_hlines}

The optimal fit to the H$\alpha$ line is shown in Figure~\ref{fig:ha}. The broadest component, with a FWHM $ = 11720_{-710}^{+560}$ km s$^{-1}$, shows signs of a mild outflow with speed $360\pm70$ km s$^{-1}$. In contrast, the narrower broad component (BC1) shows an inflow with respect to the quasar's systemic redshift of $-315 \pm 37$ km s$^{-1}$. 

We obtain a black hole mass based on the FWHM and  luminosity of the H$\alpha$ line of:
\begin{equation}
    M_{\text{BH, H-}\alpha} = 1.52 \pm 0.17 \cdot 10^9 M_\odot .
\end{equation}
For comparison, the black hole mass of J1120+0641 derived based on the rest-UV Mg~{\small{II}} emission line is \citep{Yang21}:
\begin{equation}
    M_{\text{BH, Mg~II}} = 1.35 \pm 0.04 \cdot 10^9 M_\odot .
\end{equation}
The black hole mass based on the rest-UV C~{\small{IV}} emission line is calculated by \citet{Farina22} using the scaling relations of \citet{Vestergaard06} with the corrections of \citet{Coatman17}:
\begin{equation}
    M_{\text{BH, C~IV}} = 2.40 ^{+0.06}_{-0.05} \cdot 10^9 M_\odot. 
\end{equation}
These measurements are remarkably compatible, especially considering the scaling relations on which they rest contain intrinsic scatter at the level of $0.3-0.5$ dex (i.e.~a factor of $2-3$). 

As an aside, a different black hole mass estimation can be conducted based on the C~{\small{IV}} emission line using the scaling relations of  \citet{Vestergaard06} but without the corrections of \citet{Coatman17}:
\begin{equation}
    M_{\text{BH, C~IV, no corr}} = 4.79 ^{+0.12}_{-0.11} \cdot 10^9 M_\odot. 
\end{equation}
However, the corrections of \citet{Coatman17} are most relevant for quasars presenting an asymmetric C~{\small{IV}} emission line, of which J1120+0641 is a good example. We therefore treat as fiducial the $M_{\text{BH, C~IV}}$ which implements the corrections of \citet{Coatman17}. Indeed, we find a very good agreement between the corrected $M_{\text{BH, C~IV}}$ and $M_{\text{BH, H-}\alpha}$ and $M_{\text{BH, Mg~II}}$.

\begin{figure}
\includegraphics[width=\columnwidth]{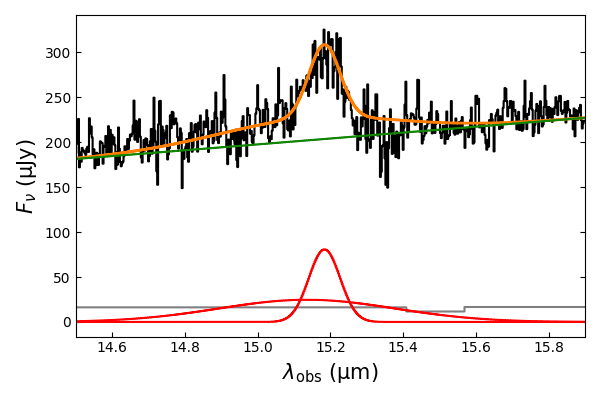}
\caption{Spectrum of the Pa-$\alpha$ broad line and the optimal model fit. The legend is the same as in Figure~\ref{fig:ha}.}
\label{fig:pa}
\end{figure}

The Eddington rate measured from the H$\alpha$ BH mass is $f = 0.9$ with measurement uncertainties of $\pm 0.1$ and systematic uncertainties from the BH mass scaling of a factor of $2$. Our measurement of J1120+0641's Eddington rate is higher but within scatter compared to previous estimates: \citet{Farina22} measure $f=0.57\pm0.01$ based on the C~{\small{IV}} emission line including only measurement uncertainties (whereas the systematic uncertainties on the C~{\small{IV}} BH mass are about $\sim 0.3-0.5$ dex).


The Paschen-$\alpha$ broad line is well described by the two-gaussian model, as can be seen in Figure~\ref{fig:pa}. However, some low-frequency spectroscopic fringing is present around the location of the line, which necessitated fixing the centre of the broadest component (BC2) in order to achieve convergence.  The centre of the broadest component is fixed to be at the same redshift as the broadest component of the H$\alpha$ line; see Section~\ref{sub:meth_lines} for details. 

With this constraint in place, the best-fit redshift of the narrower broad component of the Pa-$\alpha$ line (BC1) implies an inflow of $-415 \pm 150$ km s$^{-1}$, a value consistent with the core component of the H$\alpha$ line within uncertainties. It is therefore highly likely that the (core part of) two emission lines originate from the same part of the BLR, even though the FWHM of the core components is not fully consistent.

The FWHM of the broadest component of Pa-$\alpha$ is $11300_{-2800}^{+2000}$ km s$^{-1}$, fully compatible with the FWHM of the broadest component of H$\alpha$. The flux ratio of the core components (BC1) to the broadest components (BC2) is $\sim0.8$ and $\sim0.6$ for H$\alpha$ and Pa-$\alpha$ respectively. However, the fluxes in the BC1 and BC2 components are degenerate in the Pa-$\alpha$ line, such that the small difference is unlikely to be statistically significant. The measurement of the total flux of the Pa-$\alpha$ emission line is not affected by the degeneracy. We obtain a black hole mass using the relation from \citet{Kim10} (see Section~\ref{sub:meth_lines}):
\begin{equation}
M_{\text{BH, Pa-}\alpha} = 9.6^{+2.7}_{-1.9} \cdot 10^8 M_\odot.
\end{equation}
The resulting mass is consistent with the estimates based on the H$\alpha$ and Mg~{\small{II}} emission lines within the intrinsic scatter of the respective scaling relations. The mass is however inconsistent with the estimate based on C~{\small{IV}}. 

\begin{figure}[t]
\includegraphics[width=\columnwidth]{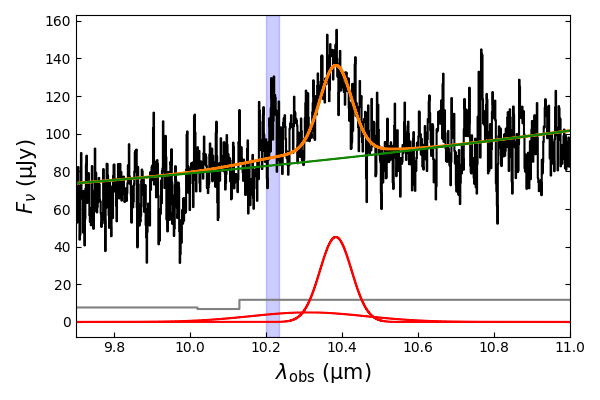}
\caption{Spectrum of the Pa-$\beta$ broad line and the optimal model fit. The legend is the same as in Figure~\ref{fig:ha}. The area in blue shows a fringing residual which skews the fit and is therefore masked.}
\label{fig:pb}
\end{figure}

We find that the Paschen-$\beta$ emission line is more challenging to fit with two gaussian components. The core component (BC1) has a relatively high SNR, and its FWHM and redshift are consistent with the properties of the core components of H$\alpha$ and Pa-$\alpha$; suggesting again an origin from a common BLR. However, we encountered significant difficulties in constraining Pa-$\beta$'s broader component, if it is detected at all. As can be seen in Figure~\ref{fig:pb}, the spectrum surrounding the Pa-$\beta$ emission line contains many fringing residuals giving rise to spurious and fairly narrow features. Despite masking a feature which otherwise causes the fit to fail to converge, the constraints we obtain are still rather poor. The optimal FWHM of BC2, $10340_{-2500}^{+1800}$ km s$^{-1}$, reflects this uncertainty while still being fully consistent with the widths of H$\alpha$'s and Pa-$\alpha$'s broadest components. Meanwhile, the shift of the broadest component is almost certainly impacted by the presence of spurious spectral features, and cannot be used to draw conclusions on the line's BLR of origin. In conclusion, the lower SNR of the line and/or the presence of spectroscopic fringing do not enable us to confidently characterise the line's broadest component. The line's total flux is relatively little impacted by these considerations since the underlying continuum is very well constrained. We obtain the black hole mass estimate based on Pa-$\beta$ using the scaling relation of \citet{Kim15} (see Section~\ref{sub:meth_lines}):
\begin{equation}
M_{\text{BH, Pa-}\beta} = 9.2^{+2.0}_{-1.6} \cdot 10^8 M_\odot.
\end{equation}
Again, the mass estimate based on Pa-$\beta$ is consistent with all other measurements obtained so far within the relation's intrinsic scatter -- with the exception of the estimate based on the C~{\small{IV}} line. Our observations therefore confirm suspicions that C~{\small{IV}} is amongst the least reliable tracers of black hole mass due to its highly-ionised and scattered nature (e.g.~\citealt{Coatman17}).

\begin{figure}[t!]
\includegraphics[width=\columnwidth]{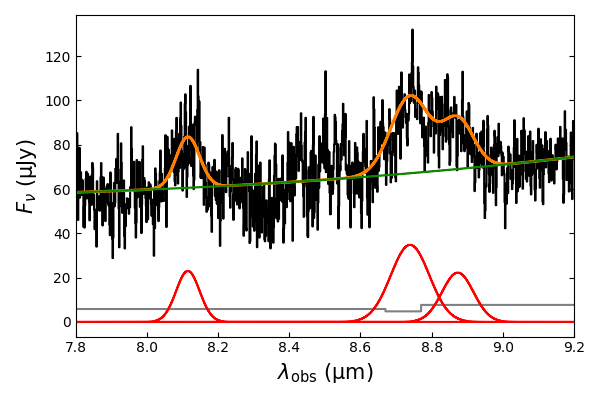}
\caption{Spectrum of the Pa-$\delta$, He~{\small{I}} and Pa-$\gamma$ broad lines (in order of increasing wavelength) and the optimal model fit. The legend is the same as in Figure~\ref{fig:ha}.}
\label{fig:pg}
\end{figure}

Before calculating the BH masses based on the combined rest-infrared properties of J1120+0641 following Equation~\ref{eq:pa}, we first discuss the usability of the remaining Pachen-series lines beyond Pa-$\alpha$ and Pa-$\beta$.

\subsection{Other Emission Lines}

The Pa-$\delta$, He~{\small{I}} and Pa-$\gamma$ broad emission lines found in our spectrum are shown in Figure~\ref{fig:pg}. The latter two lines are blended, which significantly hinders the determination of their respective fluxes and kinematics. We therefore model them with one gaussian component each, and nevertheless find many of their fit parameters to be degenerate. Based on our best-fit decomposition, the He~{\small{I}} lines appears to contain about twice the flux of the Pa-$\gamma$ line. Both lines are quite broad, with FWHMs of $4370^{+410}_{-350}$ km s$^{-1}$ and $3520_{-390}^{+440}$ km s$^{-1}$, respectively. However, the fact Pa-$\gamma$'s FWHM exceeds the width of the core components of the previous $3$ hydrogen transitions could be due to the presence of an additional broader component, which we do not have sufficient SNR to detect. Alternatively, the blended He~{\small{I}} line could also contain significant wings which contaminate the fit over Pa-$\gamma$.

Pa-$\delta$, in contrast, is not contaminated by blending, but its lower SNR of $\lesssim 2$ per pixel only warrants a single gaussian component (leftmost line in Fig.~\ref{fig:pg}). The width of the line is well captured and fully consistent with the widths of the core components of H$\alpha$, Pa-$\alpha$, and Pa-$\beta$. This further suggests that the narrower components of all detected hydrogen transitions originate from the same BLR, and that the anomalously large measurement in Pa-$\gamma$ is due to blending contamination.

Since the Pa-$\delta$ line provides a well-constrained FWHM without contamination by fringing, we opt to use it in addition to Pa-$\alpha$ and Pa-$\beta$ in our nominal Paschen-series-based BH mass estimate. Following Equation~\ref{eq:pa}, we first calculate the continuum luminosity at $1\mu$m. Using only the continuum model (accretion disk power-law and hot dust black-body), we obtain $\nu L_{1 \mu\text{m}} = 1.69 \pm 0.10 \cdot 10^{45}$ erg s$^{-1}$. To obtain the optimal FWHM$_{\text{H}}$, we combine the total line widths of Pa-$\alpha$, Pa-$\beta$ and Pa-$\delta$ and obtain a FWHM$_{\text{H}} = 2780 \pm 110$ km s$^{-1}$. Finally, we obtain:
\begin{equation}
    M_{\text{BH, IR}} = 3.7 \pm 0.4 \cdot 10^{8} M_\odot .
\end{equation}
While this BH mass is lower than measured via the other broad lines (Section \ref{sub:res_hlines}), it remains fully consistent when considering that the relation in Equation~\ref{eq:pa} contains $1$ dex of intrinsic scatter. We also calculate the BH mass based on the $3$ non-blended Paschen-series lines separately, and list all our BH mass measurements and literature values in Table 2.
\begin{figure}[t]
\includegraphics[width=\columnwidth]{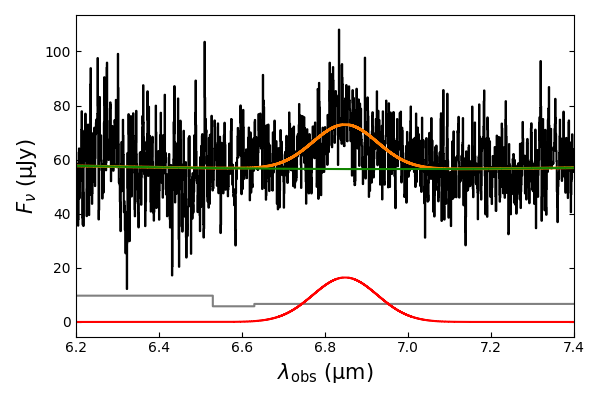}
\caption{Spectrum of the O-{\small{I}} broad line and the optimal model fit. The legend is the same as in Figure~\ref{fig:ha}.}
\label{fig:oi}
\end{figure}

The O~{\small{I}} emission line is the only detected non-hydrogen emission line which is not blended. We find O~{\small{I}} to be significantly broader than the ``core'' component of the hydrogen lines but narrower than the broadest components: FWHM$_{\text{O~I}} = 8000_{-1100}^{+800}$ km s$^{-1}$. Such a large width is more similar to the widths seen in rest-UV quasar emission lines, such as C~{\small{IV}} (e.g.~\citealt{Farina22}). We are not aware of any BH mass scaling relations calibrated to the O~{\small{I}} $0.8446 \mu$m line.

Combining multiple transitions of O~{\small{I}} has been suggested as a test of the ionisation strength in the joint O~{\small{I}}+Fe~{\small{II}} BLR, in an approach similar to that we employed for the broad hydrogen lines \citep{Kwan81,Grandi83,Matsuoka07}. However, we only have one other line available to us: the rest-UV O~{\small{I}} $1304$\AA \ line \citep{Bosman17}. We calculate the line fluxes of both transitions taking into account the slit losses in the X-Shooter spectrum, and obtain a ratio of these lines of $f_{\text{OI, }0.8446\mu\text{m}}/f_{\text{OI, }1304\text{\AA}} = 1.08 \pm 0.08$. \citet{Matsuoka07} report higher values in $6$ quasars at $z<1.5$, ranging from $1.5$ to $4.7$. However, the comparison sample is too small to draw conclusion on the possible evolution of O~{\small{I}}. In addition, the O~{\small{I}} $1304$\AA \ line is blended with a Si~{\small{II}} line at the same wavelength, from which it cannot be separated in the X-Shooter spectrum. Nevertheless, the direction of deviation would suggest a slightly lower ionisation parameter $U$ and/or lower O~{\small{I}} BLR density in J1120+0641 compared to the $6$ quasars at $z<1.5$.

\subsection{BLR density and ionisation parameter}

\begin{figure}[t]
\includegraphics[width=\columnwidth]{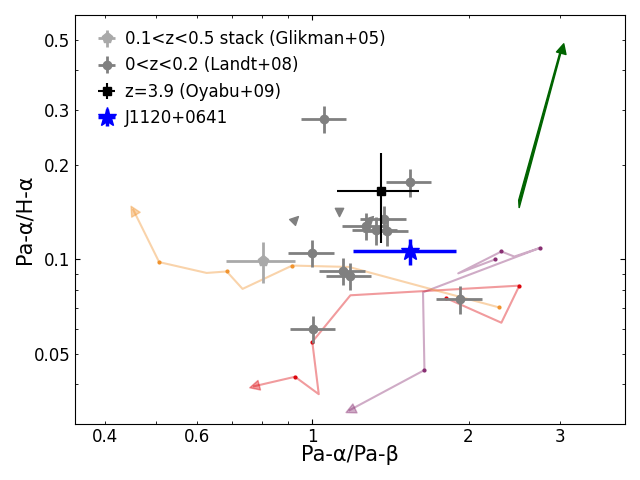}
\caption{Observed ratios of the Pa-$\alpha$, Pa-$\beta$ and H$\alpha$ emission line strengths (data points) and outputs from our {\textsc{Cloudy}} models (coloured tracks). Arrows at the end of the tracks indicate the direction of increasing ionisation parameter $U$, with dots indicating $\log U=-7, -5, -3, -1$. The tracks correspond to different BLR densities of $\log N_H /\text{cm}^{-3} = 10$ (purple, rightmost track), $12$ (red, middle track), and $14$ (yellow, leftmost track). The arrow in the top-right corner indicates the direction of extinction; its length corresponds to E(B$-$V) $=0.5$.}
\label{fig:ratio}
\end{figure}

The results of our modeling of the hydrogen BLR with \textsc{Cloudy} are shown in Figure~\ref{fig:ratio}, together with measurements from J1120+0641 and comparison samples at low redshift. Based on interpolating between outputs, we find that the ratios of H$\alpha$, Pa-$\alpha$ and Pa-$\beta$ are similar to $0<z<4$ luminous quasars, and the lines are well modeled with a BLR density of $\log N_H / \text{cm}^{-3} \geq 12$ and an ionisation parameter $-7<\log U<-4$. Very little extinction of the BLR, of the order E(B$-$V) $\lesssim 0.1$, is required to explain the observed emission line ratios. Compared to the \citet{Landt08} sample of $12$ quasars at $0<z<0.2$ for which all three emission lines are detected, J1120+0641 seems to have a somewhat low ionisation parameter $U$ (although not the lowest) as indicated by a slightly high value of Pa-$\alpha$/Pa-$\beta = 1.54\pm0.34$ but a normal value of Pa-$\alpha$/H-$\alpha = 0.106 \pm 0.010$. As also seen in many $z<4$ quasars (Fig.~\ref{fig:ratio}, see also \citealt{Soifer04}), J1120+0641's hydrogen broad emission lines are incompatible with pure case B recombination, which predicts Pa-$\alpha$/Pa-$\beta = 2$ and Pa-$\alpha$/H-$\alpha = 0.1$ \citep{Osterbrock89}. Overall, the ratios of the Paschen lines in J1120+0641 are highly similar to those of quasars at $0<z<5$.

\section{Discussion}\label{sec:disc}

\subsection{Structure of the BLR}\label{sub:BLR}

Our observations have revealed the kinematics of the H$\alpha$ and first $4$ Paschen-series emission lines of hydrogen in unprecedented detail in a $z=7$ quasar. Perhaps surprisingly, we find that hydrogen broad lines are fully consistent with a simple model of the hydrogen BLR. The ``core'' components of the H$\alpha$ and Pa-$\beta$ lines, as well as the single visible component of Pa-$\delta$, are  consistent in their widths of $\sim2880$ km s$^{-1}$, well within their $\sim5\%$ uncertainties. The broadest component, when it is detected in H$\alpha$ and Pa-$\alpha$ and marginally detected in Pa-$\beta$, is similarly extremely consistent across lines in its FWHM of $\sim11000$ km s$^{-1}$. The velocity shifts of the core components is similarly highly consistent in its value of $\sim 320$ km s$^{-1}$ (inflowing) across H$\alpha$, Pa-$\alpha$ and Pa-$\beta$. The broadest component's shift is less well attested and can only be measured with confidence in H$\alpha$, where it stands at $-360\pm70$ km s$^{-1}$ (outflowing). The only anomalies can be accounted for by the presence of residual fringing (broadest components of Pa-$\alpha$ and Pa-$\beta$), line blending (Pa-$\gamma)$, and low SNR (Pa-$\delta$).

\begin{table*}
\caption{Summary of the BH mass estimates of J1120+0641. The ``relation scatter'' column indicates the intrinsic uncertainty coming from calibration of the scaling relations for each line, while the $M_\text{BH}$ uncertainties reflect only the $1\sigma$ measurement uncertainties.}
\begin{center}
    \begin{tabular}{l|c c l}
    Line & $M_{\text{BH}}$ & Intrinsic relation scatter & Reference \\
     & $(10^9 M_\odot)$ & (dex) & \\
    \hline
    H$\alpha$ & $1.52 \pm 0.17$ & $\sim 0.3$ & This paper\\
    Mg~{\small{II}}   & $1.35 \pm 0.04$ & $0.3-0.5$ & \citealt{Yang21} \\
    C~{\small{IV}} & $2.40^{+0.06}_{-0.05}$ & $0.3-0.5$ & \citealt{Farina22} \\
    Pa-$\alpha$ & $0.96^{+0.27}_{-0.19}$ & $\sim0.2$ & This paper\\ 
    Pa-$\beta$ & $ 0.92^{+0.20}_{-0.16} $ & $\sim0.2$ & This paper\\
    Combined rest-IR & $0.37 \pm 0.04$& $\sim1$ & This paper\\
    \hline
    \end{tabular}
    \end{center}
\end{table*}

A slightly redshifted ``core" component of the broad hydrogen emission line is also seen in H-$\beta$ in a few quasars at $z>6.5$ \citep{Yang23}, matching our observations of a $\sim300$ km s$^{-1}$ ``inflow''. The physical origin of this effect is unknown. Even though it seems likely that the ``core'' and ``broadest'' hydrogen BLR components originate in different gas, seeing an inflow in the BLR would require a complex morphology of the accretion structure. Another explanation could be that we are seeing a small ``recoil'' of the SMBH and accretion structure with respect to the host galaxy. The FWHM of the [C~{\small{II}}] line, which is a rough measurement of the stellar velocity dispersion, is FWHM$_{[\text{CII}]} = 416 \pm 39$ km s$^{-1}$ \citep{Venemans20}; larger than the measured offset. The timescales on which SMBHs settle in the centers of galaxies at $z>
7$ are not known, but SMBH binary mergers are known to occur on time scales $\sim100$ Myr after galaxy mergers and introduce ``kicks'' of order $100$s--$1000$s km s$^{-1}$ (e.g.~\citealt{Campanelli07}).

Our simple \textsc{Cloudy} model appears to be sufficient to reproduce the observed ratios of the H$\alpha$, Pa-$\alpha$ and Pa-$\beta$ lines, which otherwise do not agree with pure Case B recombination. The observed ratios are consistent with a single BLR zone with density $\log N_H/\text{cm}^{-3} \geq 12$, ionisation parameter $-7<\log \ U <-4$, and BLR extinction E(B--V)$\lesssim0.1$. These parameters, and the observed line ratios, are typical for other $0<z<4$ luminous quasars. 

In summary, we do not find peculiarities in the hydrogen BLR of J1120+0641 at $z=7.0848$. The only sign of ``evolution'' in this quasar compared to its low-$z$ counterparts remains the strength of the ionised C~{\small{IV}} outflow, which is more than double that seen on average at $z<5$ \citep{Meyer19-qso,Schindler20}. Strong highly-ionised outflows are a common feature in early quasars, but we could find no obvious anomaly in the rest-IR hydrogen BLR which would help explain or contextualise the phenomenon in J1120+0641.


\subsection{Other dust models}\label{sub:dust}

Instead of the classical assumption of hot torus dust emitting maximally optically-thick black-body radiation, we attempt to fit J1120+0641's infrared continuum with optically-thin dust emission. We use a black-body slope of $\beta=2$ into the modified black-body (Eq.~\ref{eq2}). As before, all continuum and line parameters are left free to vary simultaneously. We show the result of the optically-thin dust fit in Appendix~A.

We find that optically-thin dust provides a comparably good fit to the continuum compared to black-body radiation.
The optimal emission line parameters are not modified by more than $1\sigma$ under the optically-think assumption, except for Pa-$\alpha$'s BC2 which is broader by $2.5\sigma$.
However, the assumption of optically-thin dust results in a significantly higher dust temperature of $T_{\text{dust, opt.~thin}} = 2363_{-11}^{+14}$K. The sublimation temperature of graphitic dust, which is believed to constitute quasar tori, is no higher than $2100$K. The existence of long-lived dust at a higher temperature is therefore somewhat improbable. A more plausible scenario is that of a relatively small amount of short-lived dust being stripped from the inner edge of the torus and advected in the accretion disk outflows. Since hotter dust contributes more to the observed emission exponentially with temperature but only linearly with mass, a much smaller amount of hotter, optically-thin dust might mask the contribution from a much denser and optically thick torus.

However, the optimal accretion disk slope is significantly bluer in the optically-thin case: $\beta=-0.90\pm0.03$. A bluer slope is required in order to create a near-flat spectrum around the H$\alpha$ emission line (Figure~\ref{fig:thin}). Observations of J1120+0641 with the NIRSpec spectrograph, obtained in early 2023, should soon be able to provide a strong independent constraint on the accretion disk power-law slope and break the degeneracy between the optically-thin and optically-thick interpretations of our observations.

Finally, we attempt to fit the continuum emission without a fixed value of the black-body slope, but instead letting $\beta$ vary $\in [1.5,2]$. However, the lowest values of $\beta$ are always preferred by the data, even when the prior range is increased to $[0,2]$. The driving factor for this parameter drift is a small depression in the spectrum around $9 \lesssim \lambda_{\text{obs}} \lesssim 10 \mu$m, which can be seen in the bottom panel of Fig.~\ref{fig:fit}. Since fitting this unusual feature would require unphysical values of $\beta<0$, we conclude that the feature is most likely an instrumental effect and that we are unable to constrain the black-body slope $\beta$. In light of this issue, we are not confident in the data's ability to constrain even more complicated dust models, such as those including a range of dusts with different masses and temperatures. Excluding the suspect wavelength region from our nominal (black-body) continuum fit alters the best fit temperature by less than $1\sigma$.


\section{Conclusions}\label{sec:ccl}

We obtained a MIRI-MRS spectrum of the quasar J1120+0641 at $z=7.0848$, revealing for the first time the rest-frame infrared properties of an early quasar at $z>7$. We analysed J1120+0641's hot dust torus emission and rest-frame infrared emission lines in the hope of finding clues to the fast assembly of the central SMBH. Our results can be summarised as follows:
\begin{itemize}
    \item Emission from hot dust near the accretion disk is clearly detected. The hot dust bump is well-fit by a black-body with a temperature of $T_\text{dust} = 1413.5_{-7.4}^{+5.7}$K. Compared to quasars at $0<z<6$, J1120+0641's hot dust is comparable in temperature but slightly hotter ($1/137$ low-$z$ quasars contain hotter dust). Compared to $6<z<6.5$ quasars, the temperature is more typical ($3/15$ are hotter). We postulate a weak evolution of the dust temperature with redshift, but cannot conclude based on the small sample size and potential methodological biases. %
    \item Despite the unprecedented sensitivity and resolution of the MRS instrument, we cannot use the shape of the hot dust emission to distinguish between optically-thick and optically-thin emission. 
    \item The black hole mass of J1120+0641 measured from the Balmer H$\alpha$ emission line as well as from the Paschen-series lines are consistent with the mass measured from the rest-frame UV Mg~{\small{II}} line. The H$\alpha$-based black hole mass is $M_\text{BH} = 1.52 \pm 0.17 M_\odot$. We therefore confirm that J1120+0641's SMBH, if it grew continuously at the Eddington rate, must have started from a black hole seed $\gtrsim 3000 M_\odot$ at $z=30$.
    \item We individually fit the H$\alpha$ and Pa-$\alpha$ through Pa-$\delta$ broad emission lines, and find that the kinematics of the hydrogen broad lines are consistent with originating in a common BLR. The broad hydrogen lines are characterised by a broad core of width $\sim2880$ km s$^{-1}$ inflowing at $\sim320$ km s$^{-1}$ and a broader component of width $\sim11000$ km s$^{-1}$ outflowing at around $360$ km s$^{-1}$. These kinematics are consistent with those seen in the H-$\beta$ line in other $z>6.5$ quasars.
    \item The ratios of the emission lines H$\alpha$, Pa-$\alpha$ and Pa-$\beta$ are fully compatible with similarly-luminous quasars at $0<z<4$. Using a simple \textsc{Cloudy} model, we find that the hydrogen lines are consistent with originating in a common BLR with density $\log N_H/\text{cm}^{-3} \geq 12$, ionisation parameter $-7<\log \ U <-4$, and BLR extinction E(B--V)$\lesssim0.1$. These parameters are typically good fits for lower-redshift quasars as well.
\end{itemize}

In conclusion, both the hot torus dust and hydrogen BLR structures of J1120+0641 are indistinguishable from quasars at later times. 
Our observations demonstrate the complex structures of the dusty torus and the BLR can establish themselves around a SMBH less than $760$ million years after the Big Bang. This fast evolution seems to mimic the fast metal enrichment seen in early quasars, which reaches present-day values already by $z=6$. 

We find no evidence of obscuration effects which would bias measurements of the black hole mass and of the quasar's accretion rate based on rest-frame UV diagnostics. There is no evidence in either the dusty torus nor the hydrogen BLR that J1120+0641 recently underwent a highly-obscured phase, nor any other signs of recent super-Eddington accretion activity. While we find a moderately elevated torus dust temperature, it is not yet statistically significant based on the small number of $z>6.5$ quasars for which hot dust has been observed. Further observation of early quasars with MIRI-MRS will be crucial to confirm this trend. 

\section*{Acknowledgements}

S.E.I.B., F.W.~and L.B.~acknowledge funding from the European Research Council (ERC) under the European Union’s Horizon 2020 research and innovation programme (grant agreement no.~740246 ``Cosmic Gas''). 
L.C.~acknowledges financial support from Comunidad de Madrid under Atracci\'{o}n de Talento grant 2018-T2/TIC-11612.
The Cosmic Dawn Center (DAWN) is funded by the Danish National Research Foundation under grant No.~140.

J.A-M., L.Colina , A.L.~acknowledge support by grant PIB2021-127718NB-100 
funded by MCIN/AEI/10.13039/501100011033. A.A-H.~is supported by grant PID2021-124665NB-I00 from the Spanish Ministry of Science and Innovation/State Agency of Research MCIN/AEI/10.13039/501100011033 and by ``ERDF A way of making Europe''. 
A.B.~\& G.Ö. acknowledge support from the Swedish National Space Administration (SNSA).O.I.~acknowledges the funding of the French Agence Nationale de la Recherche for the project iMAGE (grant ANR-22-CE31-0007), J.H. and D.L.~were supported by a VILLUM FONDEN Investigator grant (project number 16599). K.I.C. acknowledges funding from the Netherlands Research School for Astronomy (NOVA) and the Dutch Research Council (NWO) through the award of the Vici Grant VI.C.212.036. 
T.P.R.~would like to acknowledge support from the ERC under advanced grant 743029 (EASY). 
A.E.~and F.P.~acknowledge support through the German Space Agency DLR 50OS1501 and DLR 50OS2001 from 2015 to 2023. 

The work presented is the effort of the entire MIRI team and the enthusiasm within the MIRI partnership is a significant factor in its success. MIRI draws on the scientific and technical expertise of the following organisations: Ames Research Center, USA; Airbus Defence and Space, UK; CEA-Irfu, Saclay, France; Centre Spatial de Liége, Belgium; Consejo Superior de Investigaciones Científicas, Spain; Carl Zeiss Optronics, Germany; Chalmers University of Technology, Sweden; Danish Space Research Institute, Denmark; Dublin Institute for Advanced Studies, Ireland; European Space Agency, Netherlands; ETCA, Belgium; ETH Zurich, Switzerland; Goddard Space Flight Center, USA; Institute d'Astrophysique Spatiale, France; Instituto Nacional de Técnica Aeroespacial, Spain; Institute for Astronomy, Edinburgh, UK; Jet Propulsion Laboratory, USA; Laboratoire d'Astrophysique de Marseille (LAM), France; Leiden University, Netherlands; Lockheed Advanced Technology Center (USA); NOVA Opt-IR group at Dwingeloo, Netherlands; Northrop Grumman, USA; Max-Planck Institut für Astronomie (MPIA), Heidelberg, Germany; Laboratoire d’Etudes Spatiales et d'Instrumentation en Astrophysique (LESIA), France; Paul Scherrer Institut, Switzerland; Raytheon Vision Systems, USA; RUAG Aerospace, Switzerland; Rutherford Appleton Laboratory (RAL Space), UK; Space Telescope Science Institute, USA; Toegepast- Natuurwetenschappelijk Onderzoek (TNO-TPD), Netherlands; UK Astronomy Technology Centre, UK; University College London, UK; University of Amsterdam, Netherlands; University of Arizona, USA; University of Cardiff, UK; University of Cologne, Germany; University of Ghent; University of Groningen, Netherlands; University of Leicester, UK; University of Leuven, Belgium; University of Stockholm, Sweden; Utah State University, USA. A portion of this work was carried out at the Jet Propulsion Laboratory, California Institute of Technology, under a contract with the National Aeronautics and Space Administration. We would like to thank the following National and International Funding Agencies for their support of the MIRI development: NASA; ESA; Belgian Science Policy Office; Centre Nationale D'Etudes Spatiales (CNES); Danish National Space Centre; Deutsches Zentrum fur Luft-und Raumfahrt (DLR); Enterprise Ireland; Ministerio De Econom\'ia y Competitividad; Netherlands Research School for Astronomy (NOVA); Netherlands Organisation for Scientific Research (NWO); Science and Technology Facilities Council; Swiss Space Office; Swedish National Space Board; UK Space Agency. 

This work is based on observations made with the NASA/ESA/CSA JWST. Some data were obtained from the Mikulski Archive for Space Telescopes at the Space Telescope Science Institute, which is operated by the Association of Universities for Research in Astronomy, Inc., under NASA contract NAS 5-03127 for \textit{JWST}; and from the \href{https://jwst.esac.esa.int/archive/}{European \textit{JWST} archive (e\textit{JWST})} operated by the ESDC.

\appendix

\section{Optically-thin dust}

\begin{figure*}
\includegraphics[width=\textwidth]{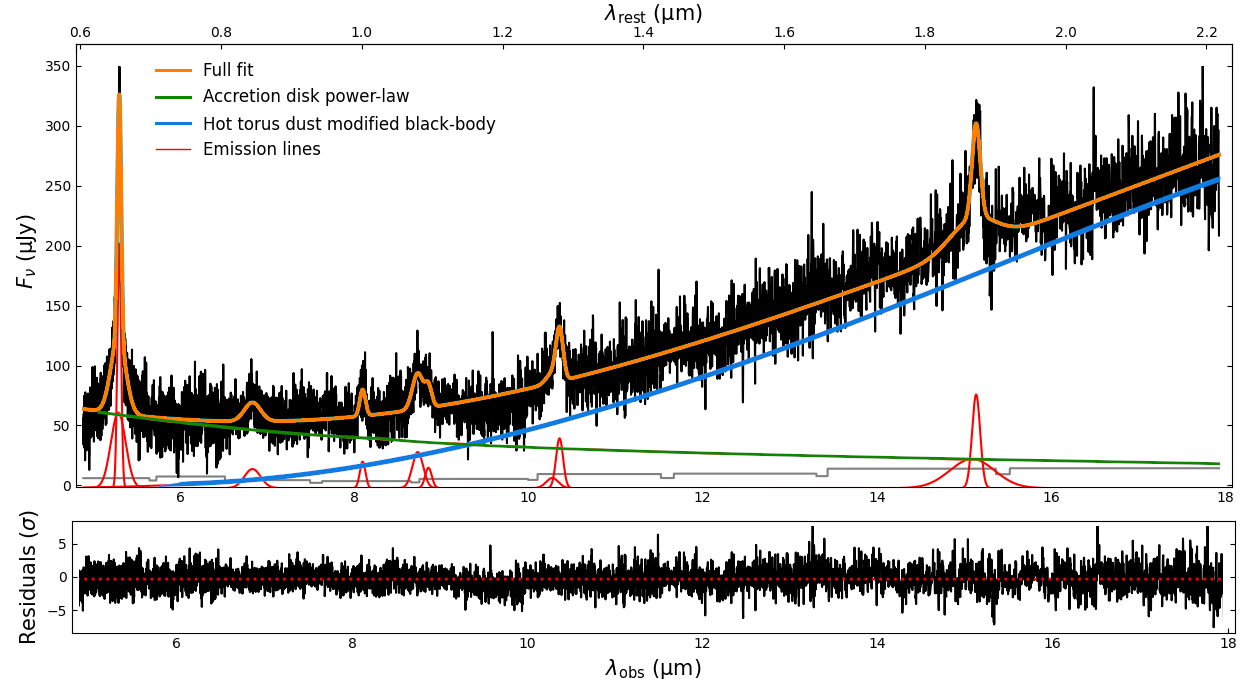}
\caption{Same as Figure~\ref{fig:fit}, but using a modified black-body with $\beta=2$ to model the hot dust continuum.}
\label{fig:thin}
\end{figure*}

Figure~\ref{fig:thin} shows the result of fitting J1120+0641's infrared emission under the assumption of optically-thin torus dust (see Section~\ref{sub:dust}).

\bibliography{bibliography}{}
\bibliographystyle{aasjournal}

\end{document}